\documentclass[aps,prl,reprint,groupedaddress,showpacs]{revtex4-1}
\usepackage{graphicx}
\usepackage{amsmath}
\usepackage{hyperref}
\pdfoutput=1

\hypersetup{
    colorlinks=true,
    linkcolor=blue,
    citecolor=blue,
    filecolor=blue,
    urlcolor=blue,
}

\begin{document}

\textbf{Comment on ``Adiabatic Quantum Algorithm for Search Engine Ranking''}

\

In their Letter \cite{QPR}, Garnerone \textit{et al.} claim that an adiabatic quantum algorithm can extract information
 about a PageRank vector with either a polynomial or exponential reduction in time resources over the classical algorithm with comparable space resources.
Here we argue that the quantum algorithm offers no obvious advantage over the classical algorithm
 and fails to preserve the fundamental stability property of the classical PageRank algorithm.

The minimum space requirements of the algorithm are set by its input.
Representation of the World Wide Web (WWW) graph requires $\tilde{\mathcal{O}}(n+m)$ space for $n$ vertices and $m$ edges,
 whether it be in classical memory, a physical network realizing the graph, or couplings between qubits within an adiabatic quantum computer (AQC).
Relative to this unavoidable cost, there is sufficient space for up to $p = \tilde{\mathcal{O}}(n+m)$ classical processors without altering spatial complexity.
In contrast, the Letter assumes $p = \mathcal{O}(1)$.

Parallelization significantly reduces the classical time complexity.
Multiplication of a vector with the Google matrix reduces to $\tilde{\mathcal{O}}(\log n)$ time
 with high-degree vertices as a limiting factor \cite{parallel_algebra}.
PageRank vector calculation with the power method reduces to $\tilde{\mathcal{O}}(\log n \log\epsilon^{-1} / \log \alpha^{-1})$ time
 for a Google matrix, $\mathbf{G} = \alpha \mathbf{P}^T + (1-\alpha) n^{-1} \mathbf{1} \mathbf{1}^T$.
The time complexity is independent of $\mathbf{P}$
 and only weakly dependent on $\alpha$, which is essential for stability.

Runtime of the quantum PageRank algorithm depends on the choice of $\mathbf{P}$.
The Letter examines average runtime and does not determine the worst-case runtime
 that sets the formal big-$\mathcal{O}$ time complexity in terms of $\{n,\epsilon, \alpha\}$.
Over $s \in [0,1]$, the AQC PageRank Hamiltonian is
\begin{equation}
 \mathbf{H}(s) = s(\mathbf{I} - \mathbf{G})^T(\mathbf{I} - \mathbf{G}) + (1-s) (\mathbf{I} - n^{-1} \mathbf{1} \mathbf{1}^T).
\end{equation}
To bound the AQC runtime, we need an upper bound on
\begin{equation}
 \Lambda = \left\|  (\mathbf{I} - \mathbf{G})^T(\mathbf{I} - \mathbf{G}) - \mathbf{I} + n^{-1} \mathbf{1} \mathbf{1}^T \right\|_2
\end{equation}
 and a lower bound on the minimum eigenvalue gap,
\begin{equation}
 \delta = \min_{s \in [0,1]} \left( \lambda_2 [ \mathbf{H}(s) ] - \lambda_1 [ \mathbf{H}(s) ] \right) .
\end{equation}
With the triangle inequality and $\|\mathbf{G}\|_2 = 1$, $\Lambda \le 5$.
For the best case ($a\propto \log \epsilon^{-1}$, $b=1$ in Eq. (5) of the Letter),
 the quantum algorithm requires $\tilde{\mathcal{O}}(\delta^{-1} \log \epsilon^{-1})$ time.

We determine the worst-case $\delta$ by minimizing over all $\mathbf{P}$ for each $\{\alpha,n\}$.
It is difficult to prove a minimum, but numerical experiments suggest that $\delta$ is minimized by
\begin{align}
 \mathbf{P} = \bordermatrix{~ & _1 & _{n-2} & _1 \cr
                  _{\lfloor n/2 \rfloor} & \mathbf{1} & 0 & 0 \cr
                     _{\lceil n/2 \rceil} & 0 & 0 & \mathbf{1} \cr} ,
\end{align}
 with $\delta^{-1} \approx 0.5 (1-\alpha)^{-2} n$.
Similar behavior is observed in WWW-like graphs, $\delta^{-1} \propto n^{0.85}$,
 for typical instances \cite{QPR2}.
Thus the time complexity of the quantum PageRank algorithm is at least $\tilde{\mathcal{O}}(n \log \epsilon^{-1} (1-\alpha)^{-2} )$.
Bounds on runtime diverge as $\alpha \rightarrow 1$ in both algorithms
 but with a relatively weaker quantum bound by a factor of $n$.

We have discussed the complexity of PageRank vector preparation
 but not the extraction of information.
The Letter considers two cases, extraction of one element and the inner product between two PageRank vectors.
Both operations require $\mathcal{O}(\log n)$ time on a parallel classical computer,
 and thus do not alter the conclusion that the existing classical PageRank algorithm
 implemented on a parallel computer is superior to the quantum adiabatic PageRank algorithm proposed in the Letter.

The results of this Comment also pertain to two recent papers proposing quantum algorithms for solving $n$-by-$n$ sparse linear systems \cite{QLS,QLS2}
 that similarly do not consider the possibility of a parallel classical computer with $\Omega(n)$ processors.
In either algorithm, the classical time complexity depends on a matrix property, $(1-\alpha)^{-1}$ here and the condition number $\kappa$ for linear systems,
 that is amplified by the quantum algorithm, $(\log \alpha^{-1})^{-1} \Rightarrow n (1-\alpha)^{-2}$ here and $\sqrt{\kappa} \Rightarrow \kappa$
 for linear systems.
Without structure in the solution vector, the classical space complexity is limited by its $\mathcal{O}(n)$ independent vector elements.
Quantum algorithms store this vector in $\mathcal{O}(\log n)$ space, and their
 total space complexity depends on the information content of the matrix rather than its dimension.
For a Google matrix, the information content is proportional to the dimension, and the quantum algorithm provides no
 advantage over the classical algorithm.
In other cases, such as a $k$-local qubit Hamiltonian for $k \ll n$, a quantum algorithm is able to achieve
 a lower spatial complexity with space-efficient procedural matrix generation.

\begin{acknowledgments}
This work was supported by the Laboratory Directed Research and Development program at Sandia National Laboratories.
Sandia National Laboratories is a multi-program laboratory managed and  
operated by Sandia Corporation, a wholly owned subsidiary of Lockheed 
Martin Corporation, for the U.S. Department of Energy's National  
Nuclear Security Administration under contract DE-AC04-94AL85000.
\end{acknowledgments}

Jonathan E. Moussa \\
\footnotesize Sandia National Laboratories, Albuquerque, NM 87185, USA \\
godotalgorithm@gmail.com

\end{document}